\documentclass{sf2a-conf2012}
\usepackage{graphicx}
\usepackage{hyperref}
\usepackage{amssymb}
\usepackage[]{natbib}  

\def\BibTeX{{\rm B\kern-.05em{\sc i\kern-.025em b}\kern-.08em
    T\kern-.1667em\lower.7ex\hbox{E}\kern-.125emX}}
\bibpunct{(}{)}{;}{a}{}{,}  

\begin{document}

\TitreGlobal{SF2A 2012}


\title{Planetary Nebulae: getting closer to an unbiased binary fraction}

\runningtitle{Planetary Nebulae: getting closer to an unbiased binary fraction}

\author{D. Douchin$^{1,4}$}

\author{O. De Marco}\address{Astronomy, Astrophysics and Astrophotonics Research Center and Department of Physics and Astronomy, Macquarie University, Sydney, NSW 2122, Australia}

\author{G.~H. Jacoby}\address{GMTO Corp, Pasadena, CA 91101, USA}
\author{T.~C. Hillwig}\address{Dept. of Physics and Astronomy, Valparaiso University, Valparaiso, IN 46383, USA}

\author{D.~J. Frew$^1$}
\author{I. Bojicic$^1$}
\author{G.~Jasniewicz}\address{Laboratoire Univers et Particules de Montpellier (LUPM), UMR 5299 - CC72, Universit\'e Montpellier 2, Place Eug\`ene Bataillon, 34 095 Montpellier Cedex 5, France}
\author{Q.~A. Parker$^1$}




\setcounter{page}{237}


\maketitle


\begin{abstract}
Why 80\% of planetary nebulae are not spherical is not yet understood. The \emph{Binary Hypothesis} states that a companion to the progenitor of the central star of a planetary nebula is required to shape the nebula and even for a planetary nebula to be formed at all. A way to test this hypothesis is to estimate the binary fraction of central stars of planetary nebula and to compare it with the main sequence population. Preliminary results from photometric variability and infrared excess techniques indicate that the  binary fraction of central stars of planetary nebulae is higher than that of the putative main sequence progenitor population, implying that PNe could be preferentially formed via a binary channel. This article briefly reviews these results and future studies aiming to refine the binary fraction.
\end{abstract}



\section{Introduction}

Planetary nebulae (PNe) are presumed to be ejected by all $\simeq$~1-8~M$_{\odot}$ stars, however when the observed PN population is compared with the theoretically expected population, a discrepancy in the number of objects appears: there are less PNe  in the Galaxy than there should be \citep{MoeAndDeMarco-2006}. This could imply that only a subset of the parent population is actually forming PNe. This subset could be the binary progenitors \citep{MoeAndDeMarco-2012}. Indeed, more than 80\%  of PNe are non-spherical \citep{ParkerEtAl-2006,JacobyEtAl-2010}, showing structures such as lobes  and jets that give an axisymmetric, point-symmetric or asymmetric shape to the nebula. The hypothesis traditionally used to account for these shapes has been the action of a magnetic field of the AGB star during the super wind phase upon the gas being ejected. However, this hypothesis has been disputed by \citet{Soker-2006} and \citet{NordhausBlackmanAndFrank-2007}, who showed that the magnetic field cannot be sustained for long enough on a whole-star scale due to the coupling between the magnetic field and the stellar rotation. Another  hypothesis to account for the non-spherical shapes of PNe is the presence of a companion (e.g. \citealt{Soker-1997}). The hypothesis according to which a companion is required to shape an axisymmetric PN has been dubbed the \emph{Binary Hypothesis} \citep{DeMarco-2009}. To test it, it is necessary to estimate the  binary fraction of central stars of planetary nebulae (CSPNe). If the observed binary fraction of CSPNe population is superior to that of the putative parent population (the main sequence (MS) stars with mass $\simeq$~1-8~M$_{\odot}$, \citealt{MoeAndDeMarco-2006}), this indicates that PNe are preferentially  a binary phenomenon (see \citet{DeMarco-2009} for a detailed review). This paper describes briefly  current  efforts aimed at estimating an unbiased binary fraction of CSPN.


\section{The binary fraction obtained using photometric variability}

Photometric variability of a binary CSPN can be due to an irradiation effect from the hot CSPN onto the companion, tidal deformations and eclipses \citep{Bond-2000,MiszalskiEtAl-2009a}. The advantage of detecting bniaries using photometric variability is that it simply requires repeated observations of targets in average observing weather conditions and from the ground. For this reason, it is a reliable method and provides constantly new results (e.g. \citealt{HillwigEtAl-2010}). The main drawback of this method is that it is biased to small separations as irradiation effect, tidal deformations and eclipses all increase in intensity or frequency with decreasing separations, therefore it only gives access to the short period binary fraction.\\
\citet{Bond-2000} and \citet{MiszalskiEtAl-2009a} already estimated close binary fractions (P$~\lesssim~$3 days) of CSPNe of  10-15\% and 12-21\% respectively. Although these fractions are lower limits, comparing them with the  MS stars binary fraction at appropriately small separations i.e. 5-7\% (\citealt{DuquennoyAndMayor-1991,RaghavanEtAl-2010,DeMarcoEtAl-2012} reveals that more PNe are formed around binaries. \\

Hillwig et al. (these proceedings) are monitoring targets from the 2.5 kpc volume-limited sample of \citet{Frew-2008} to estimate a new close binary fraction. Although the method is similar, the sample is less biased than the previously used magnitude-limited samples and also deeper (V $<$ 21). In a similar experiment, \citet{JacobyEtAl-2012} are monitoring  5-6 CSPNe within the Kepler satellite field of view, to estimate an independent binary fraction. The sample is statistically small ; however, the CSPNe are observed with a precision never reached before ($\lesssim~$1 mmag).

\section{The  binary fraction obtained  using red and infrared excess}

The red/IR excess technique aims to detect the signature of a cool, unresolved companion by measuring the absolute photometry of the CSPN. To do so, high precision absolute photometry needing photometric weather conditions in the $B$, $V$ and $I$ or $J$ bands is required. This technique is fully described in \citet{DeMarcoEtAl-2012}.  The measured $B-V$ color is compared to the expected $B-V$ for the single CSPN temperature according to atmospheric stellar models (e.g. \citealt{RauchAndDeetjen-2003}) and allows reddening to be determined whereas the $V-I$ or $V-J$ color allows the measurement of the red/IR excess, which is the difference between the $V-I$ or $V-J$ expected for a single star at the CSPN temperature and the  measured one. If this difference is greater than the error on the photometric measurement, it is a binary detection. Since companions cooler than $\simeq$ M0-5  are faint, we need excellent photometric precision. Once a binary fraction has been estimated, it can be compared to the MS one \citep{RaghavanEtAl-2010} only after undetected systems are accounted for. Using the $J$-band allows to detect colder companions, while still not being contaminated by hot dust, although it requires a separate NIR observing run and is therefore time demanding.
\\
\citet{FrewAndParker-2007} have used the photometry from the 2MASS and DENIS NIR  surveys to determine a  binary fraction $\simeq$~54\% but the detection bias was poorly quantified. \citet{DeMarcoEtAl-2012} have used the method described above on a sample of 27 CSPNe and have found a  debiased fraction $\simeq$ 30\% from $I$-band data and ~54\% from $J$-band data of a subset of 11 CSPNe in line with \citet{FrewAndParker-2007}  J-band results. These preliminary results will be confronted by the study of an additional 23 objects for which optical absolute photometry has been acquired at the NOAO 2.1m telescope in March 2011 as well as $\simeq$ 30 objects for which $J$ and $H$-band photometry has been obtained at the AAT 4m telescope in 2011 and the ANU 2.3m telescope in 2012. These new measurements should bring the sample to a statistically significant size and considerably reduce the error bars on the binary fraction. Recent surveys including $J$-band photometry will be analysed as well to extract the IR excess of other targets from the sample of Frew (2008).

\section{Conclusion}

Estimating an unbiased binary fraction of CSPNe is crucial to understanding whether companions play a key role in shaping PNe. Photometric variability has allowed us to determine a close binary fraction of 15-20\% and is still being refined using a new, less biased sample to understand the biases inherent to the method. The red/IR excess technique has allowed us to obtain a CSPN binary fraction of 70-100\%, much larger than for the MS population. However, this number carries a large uncertainty for the moment due to the small sample size. Current studies based on optical and NIR photometry as well as the use of recent NIR surveys will double the sample size to constrain the CSPN binary fraction precisely enough to support or refute the hypothesis that PNe could emerge preferentially from binary star evolution.

\bibliographystyle{aa}  
\bibliography{bibliography} 

\end{document}